\documentclass[twoside,english]{iopart}
\usepackage{mathpazo}
\usepackage[T1]{fontenc}
\usepackage[latin9]{inputenc}
\usepackage[letterpaper]{geometry}
\usepackage{amstext}
\usepackage{graphicx}
\usepackage{esint}

\makeatletter

\providecommand{\tabularnewline}{\\}

\usepackage{iopams}
\usepackage{setstack}

\usepackage[numbers, sort&compress]{natbib}
\usepackage{url}

\makeatother

\usepackage{babel}

\begin{document}

\title[]{Detection of network inhomogeneity by total neighbor degree }

\author{{\Large Hon Wai Lau, Kwok Yip Szeto}}

\address{{\large Department of Physics, The Hong Kong University of Science
and Technology, Hong Kong, China}}

\ead{{\large lau65536@gmail.com, phszeto@ust.hk}}
\begin{abstract}
Inhomogeneity in networks can be detected by the analysis of the correlation
of the total degree of nearest neighbors. This is illustrated by two
models. The first one is a random multi-partitions network that the
Aboav Weaire law, which predicts the linear relationship between the
degree of node and the total degree of nearest neighbor, is being
extended. The second one is a preferential attachment network with
two partitions which shows scale free properties with power tail $\gamma$
within the range $2<\gamma\le3$. By plotting the total degree of
neighbor verses the degree of each node in the networks, the scattered
plot shows separable clustering as evidence for inhomogeneity in networks.
The effectiveness of this new tool for the detection of inhomogeneity
is demonstrated in real bipartite networks. By using this method,
some interesting group of nodes of semantic and WWW networks have
been found. 
\end{abstract}



\submitto{\NJP}

\maketitle

\section{Introduction}

Many abstract structures and complex systems can be conveniently described
by network. Examples include World Wide Web, power grid, food webs,
word co-occurrence and protein interaction network \cite{Albert2002-Rev,Barabasi1999-science}.
Depending on the system, some of them are naturally modeled as bipartite
networks such as movie-actor network \cite{Barabasi1999-science},
paper coauthorship \cite{Newman2001a,Newman2004}, protein interaction
\cite{metabolic-network}, sexual relationship \cite{Liljeros2001-sexual,Bearman2004-sexual},
music sharing \cite{Lambiotte2005-music} and soccer championship
\cite{Onody2004-soccer}. One common tool to analyze a bipartite network
is to project it into either set of nodes that are of interest to
the specific investigation. In the projection, edges are usually reconnected
as a complete subgraph \cite{Newman2003,Newman2004a-pnas}, possibly
with some weighting schemes \cite{Zhou2007}. However, information
is inevitably lost in any projection methods and even undesirable
such as projecting sexual network into a female only or male only
uni-partite network. Thus, some analyses have been carried out on
the bipartite network directly \cite{Evans2007,Holme2003}. In general,
a network with two partitions can have internal links and the resemblance
with the bipartite network can be measured by bipartivity \cite{Estrada2005,Holme2003}.
In real world network, different type of nodes may not be known explicitly,
so it is interesting to classify nodes into different groups \cite{Guillaume2006}.
A similar problem is community detection that focuses on classifying
nodes into different partitions by optimizing modularity \cite{Barber2007,Reichardt2006-SM-community}.

The local environment of a node can be described by some typical properties
such as degree and clustering coefficient. However, it is possible
that the clustering coefficient is close to zero for a nearly bipartite
network. In this case, the degree of the neighbors can therefore provide
significant information. The degree-neighbor degree correlation can
be characterized by the Pearson correlation coefficient \cite{Newman2002}
and it has been studied for networks with homogeneous neighbor degree
by using the Aboav Weaire (AW) law \cite{Aboav1970,Weaire1974,Aboav1980}.
The original works done by Aboav and Weaire in two-dimensional cellular
networks have shown an empirically linear relationship between the
averaged total neighbor degree and the node degree. Later, this law
has been generalized for random, scale free and some real world networks
\cite{Ma&Szeto2006}. The failure of this empirical observation thus
provides a hint to detect local inhomogeneity of networks and possibly
a method to group nodes together. 

The aim of this paper is to study the correlation between total degree
of nearest neighbors and the degree of the node itself for network
with more than one partition. Through the extension of the AW law,
we propose a simple method to classify nodes of nearly bipartite networks
and modular networks. The rest of this paper is organized as follows:
In section \ref{sec:Multipartitions-random-network}, we present a
model of random multi-partition network and derive the generalization
of the Aboav Weaire law of this network. In section \ref{sec:Preferential-attachment-network},
we study a preferential attachment network with two partitions and
discuss the implication of this model to real world network. In section
\ref{sec:Degree-Neighbor-degree-correlation}, we examine node degree
and nearest neighbor degree correlation of real world networks for
both bipartite networks and non-bipartite networks. The results suggest
an interesting finding for the network of WordNet and WWW. Finally,
a conclusion is given in section \ref{sec:Summary}.

\section{Multi-partitions random network \label{sec:Multipartitions-random-network}}

We begin our studies with a brief review of the Aboav-Weaire law.
The AW law states that for a node with degree $k$ and mean neighbor
degree $M(k)$, the averaged total neighbor degree has a linear relation
with the node degree, given by $\left\langle kM(k)\right\rangle =Ak+B$,
where $A$ and $B$ are the parameters depending on the network. For
a random network with degree distribution $\mathcal{P}(k)$, the probability
$\mathcal{Q}(k)$ of selecting one of the nearest neighbors with degree
$k$ is proportional to the total degrees of all nodes with degree
$k$, or $N\mathcal{P}(k)k$. Hence, after normalized, we get $\mathcal{Q}(k)=\mathcal{P}(k)k/\left\langle k\right\rangle $
\cite{Newman2001} and the mean degree of neighbor is $\int\left(P(k)k/\left\langle k\right\rangle \right)kdk=\left\langle k^{2}\right\rangle /\left\langle k\right\rangle $.
The Poisson degree distribution of random network gives $\left\langle k^{2}\right\rangle =\left\langle k\right\rangle ^{2}-\left\langle k\right\rangle $
and so $A=\left\langle k\right\rangle +1$ \cite{Ma&Szeto2006}. In
addition, the parameter $B$ represents the assortivity of a network
and it takes value $B=0$.

Now, we consider a random network consists of $n$ different partitions
that the inter-connections are specified between each partition. Formally,
a random $n$-partition network with a set of nodes $V$ is composite
by $n$ disjoint partition of nodes $\{V_{1},...,V_{n}\}$. For the
network with size $|V|=N$, the size of each partition $V_{i}$ can
be specified by the the fraction of nodes $r_{i}=|V_{i}|/N$ such
that $\sum_{i}r_{i}=1$. The edge between each partition can be characterized
by the probability matrix $\mathbf{P}=(p_{ij})$ that $p_{ij}$ represents
the probability that any two nodes in $V_{i}$ and $V_{j}$, respectively,
are connected. Hence, the diagonal entries represent the self-linkage
probabilities and off-diagonal entries are the cross-linkage probabilities.
For this model, the mean degree contribution from partition $V_{j}$
to $V_{i}$ is $p_{ij}r_{j}N$. By summing up all degree contribution
from different partitions, we can get the mean degree for the partition
$V_{i}$: \begin{equation}
\left\langle k\right\rangle _{i}=\left(\sum_{j=1}^{n}p_{ij}r_{j}\right)N\end{equation}
where $\left\langle \cdot\right\rangle _{i}=\int\cdot\mathcal{P}_{i}(k)dk$
is the average taken over the degree distribution of partition $V_{i}$.
Here, we consider the model with large $N$ limit such that the $N\gg p_{ij}N\gg1$,
so the probability distribution of each partition is close to a continuous
Poisson distribution $\mathcal{P}(k)$ with the mean degree $\left\langle k\right\rangle _{i}$.
Moreover, the mean degree $\left\langle k\right\rangle $ of the whole
network is given by the weighted average of the mean degree of each
partition, $\left\langle k\right\rangle =\sum_{i}r_{i}\left\langle k\right\rangle _{i}$.

\begin{figure}
\noindent \begin{centering}
\includegraphics[width=0.5\textwidth]{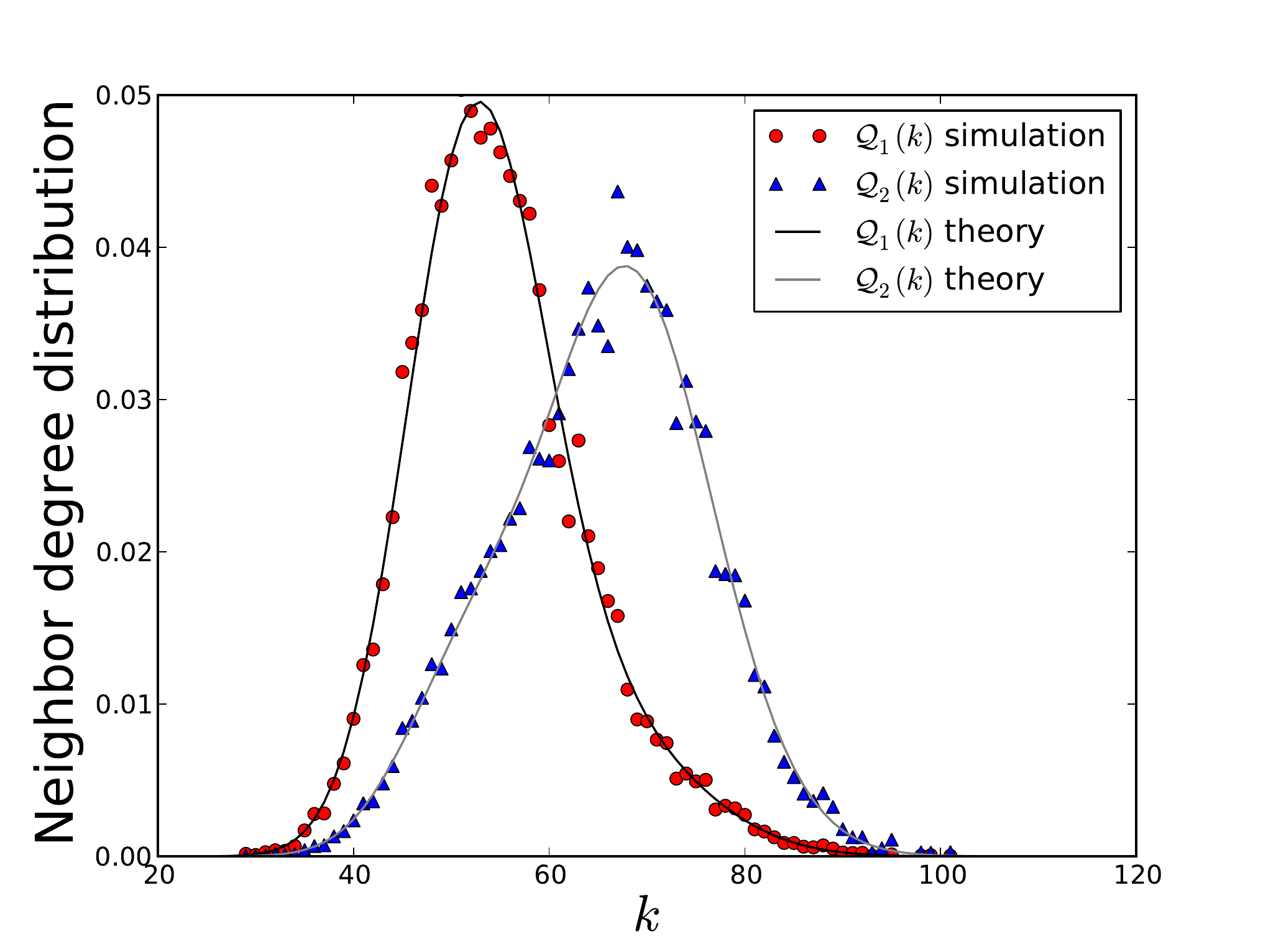}\includegraphics[width=0.5\textwidth]{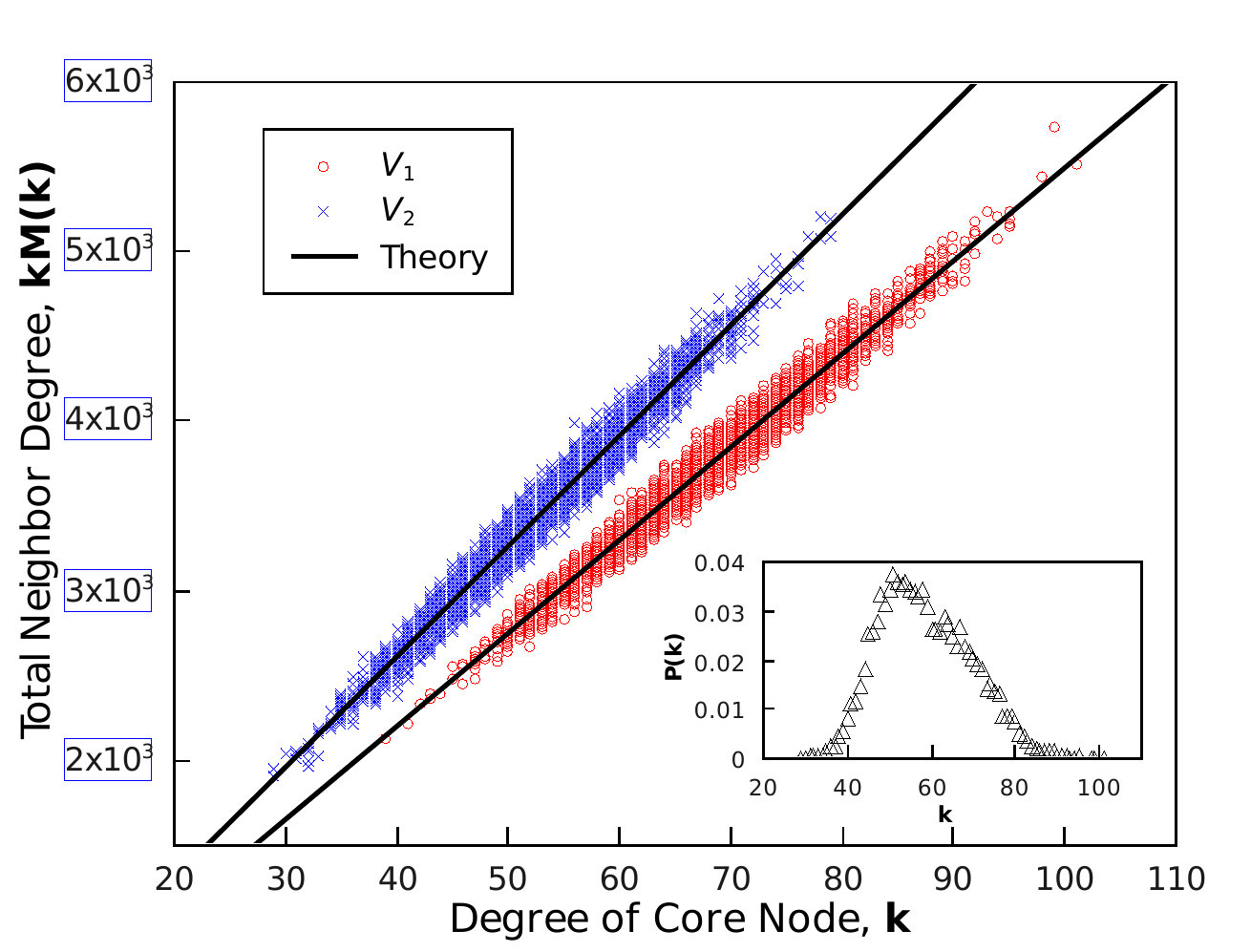}
\par\end{centering}

\caption{\label{fig:random} (a) The nearest neighbor degree distribution $\mathcal{Q}(k)$
of the random network with two partitions (b) Scattered plot of total
neighbor degree $kM(k)$ vs $k$. $N=10000$, $r_{1}=0.4$, $p_{ii}=0.002$,
$p_{12}=p_{21}=0.01$. }

\end{figure}

The nearest neighbor degree distribution $\mathcal{Q}_{i}(k)$ is
similar to the case of simple random network: a randomly selected
neighbor located in partition $V_{j}$ gives degree distribution $\mathcal{P}_{j}(k)k/\left\langle k\right\rangle _{j}$.
For a node $u$ in the partition $V_{i}$ with degree $k_{u}$, on
average, there are $k_{u}(p_{ij}r_{j}N/\left\langle k\right\rangle _{i})$
edges connected to partition $V_{j}$. Therefore, the distribution
$\mathcal{P}_{j}(k)k/\left\langle k\right\rangle _{j}$ of each partition
has to be weighted by the factor which is proportional to $p_{ij}r_{j}$.
After normalized, we have\begin{equation}
\mathcal{Q}_{i}(k)=\frac{N}{\left\langle k\right\rangle _{i}}\sum_{j}p_{ij}r_{j}\frac{\mathcal{P}_{j}(k)k}{\left\langle k\right\rangle _{j}}\label{eq:nei-deg-distri}\end{equation}
For this random network, it gives very good approximation because
there is no strong degree correlation. The result is shown in Fig.
\ref{fig:random}a with the plot of a simulation result of $\mathcal{Q}_{i}(k)$
and the predicted result using the theoretical value of $\mathcal{P}_{j}(k)$
and $\left\langle k\right\rangle _{j}$.

For this model, the nodes in the same partition have homogeneous local
environment. Hence, it is expected the linear relation of the total
neighbor degree should still be hold for each partition separately,
with the form $\left\langle kM(k)\right\rangle _{i}=A_{i}k+B_{i}$.
For this random network, $B_{i}=0$ and the mean neighbor degree for
the partition $V_{i}$ is

\begin{equation}
\left\langle M(k)\right\rangle _{i}=\frac{N}{\left\langle k\right\rangle _{i}}\sum_{j}p_{ij}r_{j}\frac{\left\langle k^{2}\right\rangle _{j}}{\left\langle k\right\rangle _{j}}\label{eq:nei-deg-slope}\end{equation}
Therefore, the slope between the total neighbor degree and node degree
is $A_{i}=\left\langle M(k)\right\rangle $. In Fig. \ref{fig:random}b,
the prediction using the above equation shows a good fit with the
simulation result. Also, from the figure, it can be observed that
the points for two partitions are separated into two clusters, while
the degree distribution of both partition are collapsed together as
shown in the inset of Fig. \ref{fig:random}b. Hence, different type
of nodes in the network can be revealed by the total neighbor degree.
Note that the linear result is not valid for the whole network because
this inhomogeneous local environment can only result in a non-linear
curve, which is the superposition of two lines in the figure. Other
than the nearly bipartite network, the clear separation is also hold
for modular network with strong internal links because there can still
have a large different in the degree between the internal nodes and
external nodes.

\section{Preferential attachment network with two partitions \label{sec:Preferential-attachment-network}}

Most networks in real world exhibit scale free behavior that the degree
distribution follows a power law at the high degree region. This property
has been studied extensively and the representative model is the BA
network introduced by Barabási and Albert \cite{Barabasi1999-science}
that is constructed by the mechanism of growth and preferential attachment.
Here, we propose a model using similar mechanism together with different
type of nodes labeled explicitly. In addition to the scale free property,
the degree of nearest neighbor of this model exhibits a rich local
behavior that the original BA model does not have.

Here, we focus on the study of two partitions network with growth
and preferential attachment for simplicity. In this model, we can
specify the ratio of node in each partition by $r_{i}$, such that
$r_{1}+r_{2}=1$, and the number of edges added at each time step
by a symmetric matrix $M=(m_{ij})$. We begin with a small network,
such as a complete bipartite network or two nodes network with one
edge. At each time step, one new node is added to the network, either
belong to the partition $V_{1}$ with the probability $r_{1}$, or
belong to the partition $V_{2}$ with probability $r_{2}$. In this
grow process, the prescribed partition size ratio $r_{1}\approx|V_{1}|/N$
can be maintained. For every newly added node located at partition
$V_{i}$, there are fixed number of edges $m_{ij}$ added between
the new node and each partition $V_{j}$ in the network. A node with
higher degree has higher chance to be connected, so for a old node
$v$ located at partition $V_{j}$, the probability of being connected
$\rho_{j}(k_{v})$ is linearly proportional to its own degree $k_{v}$.
After normalized by the total degree of its own partition:\begin{equation}
\rho_{j}(k_{v})=\frac{k_{v}}{\sum_{\alpha\in V_{j}}k_{\alpha}},\qquad\text{for }k_{v}\in V_{j}\label{eq:preferential_attachment}\end{equation}

\begin{figure}
\noindent \begin{centering}
\includegraphics[width=0.5\textwidth]{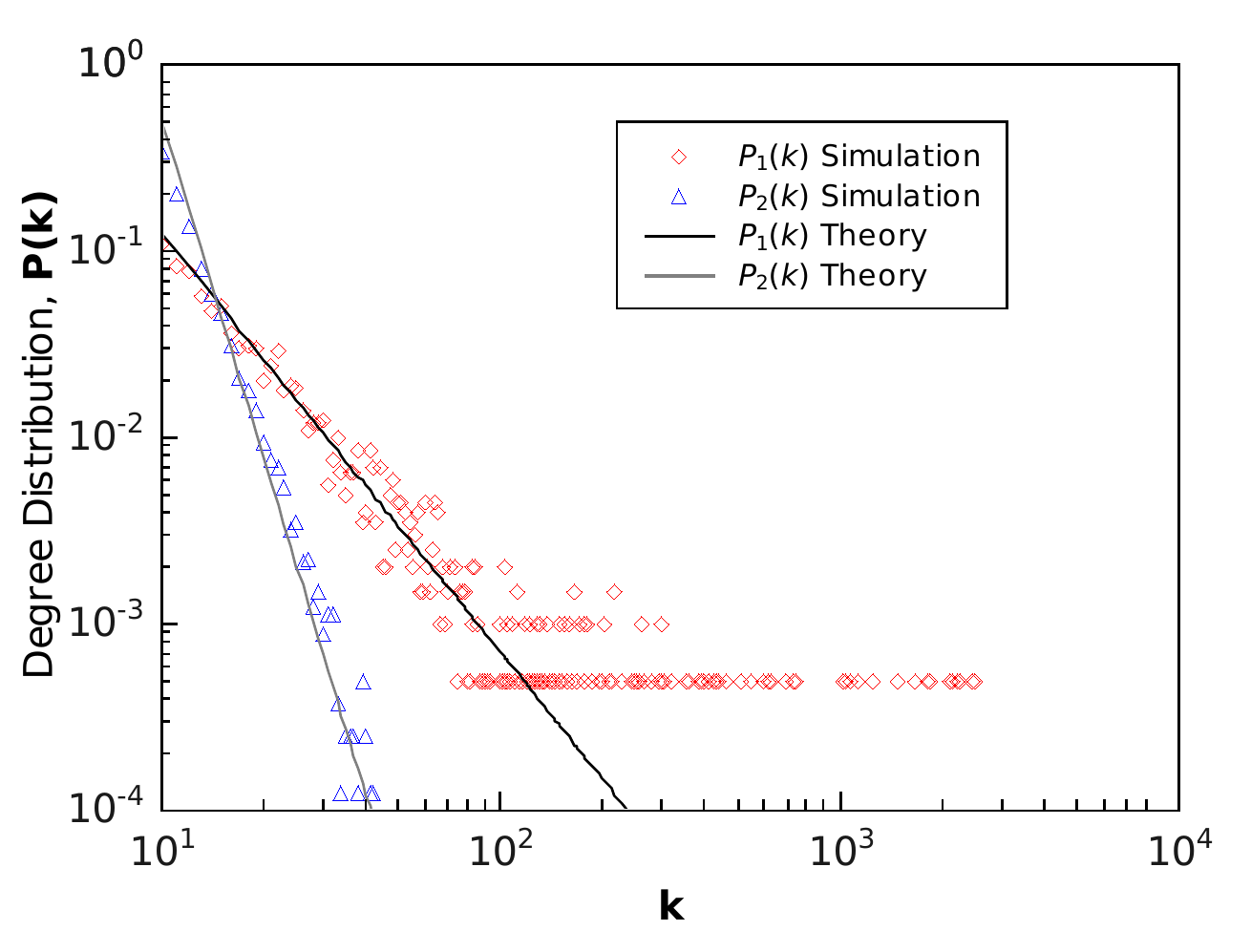}\includegraphics[width=0.5\textwidth]{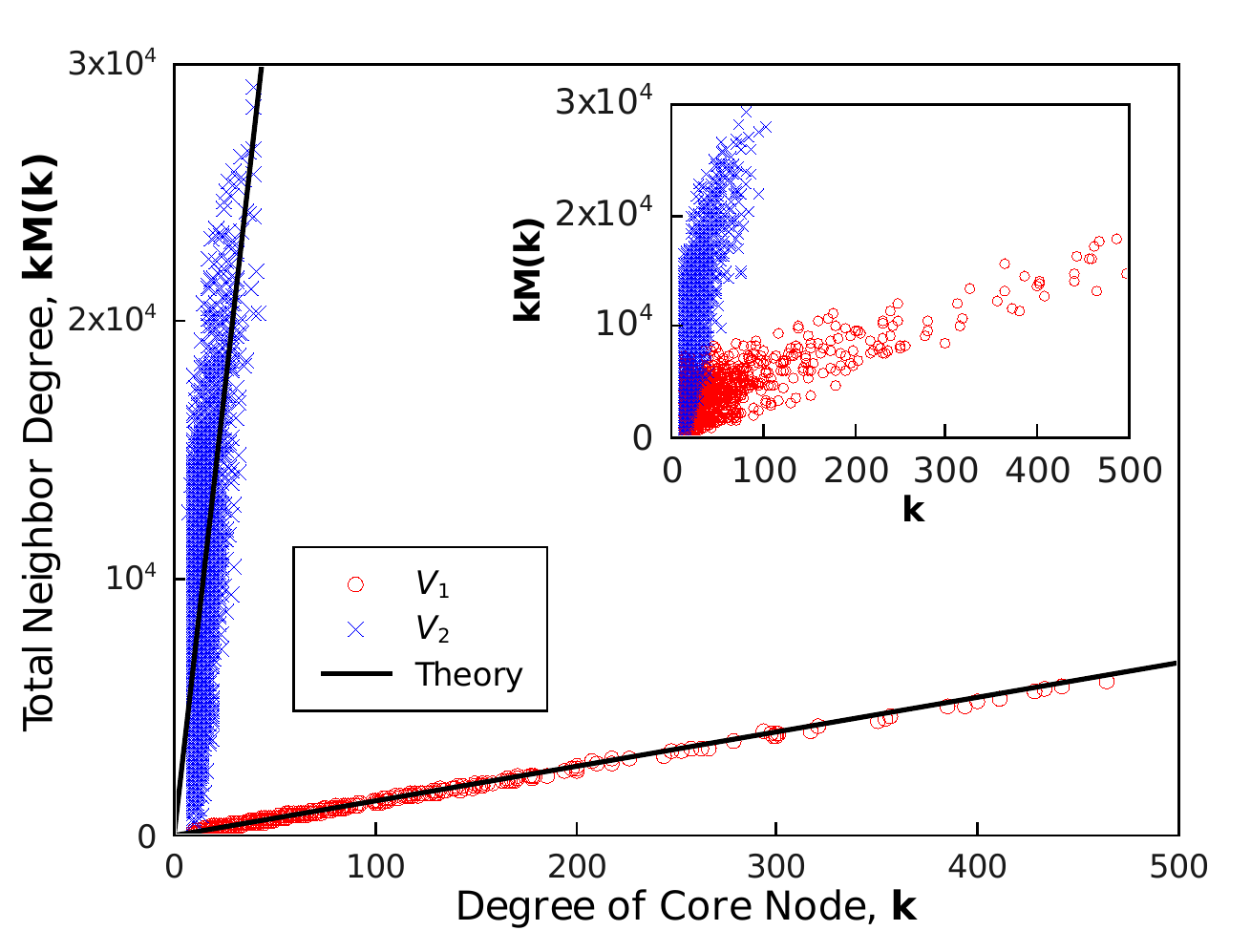}
\par\end{centering}

\caption{\label{fig:scalefree} (a) Degree distribution $\mathcal{P}_{i}(k)$
for the partitions $V_{1}$ and $V_{2}$ of the scale free model with
$N=10000$, $r_{1}=0.2$, $m_{ii}=0$, $m_{12}=10$ . (b) Scattered
plot of total neighbor degree $kM(k)$ vs $k$. (inset) The corresponding
network with some internal links $m_{ii}=3$.  }

\end{figure}

After describing the model, we now look for the evolving of the degree
of a given node $u$ in the partition $V_{1}$. Similar result for
nodes in partition $V_{2}$ can be obtained by interchange the index
1 and 2. From the point of view of an old node $u$, it has, on average,
$m_{11}\rho_{1}(k_{u})$ edges added for a new node in $V_{1}$ and
$m_{12}\rho_{1}(k_{u})$ edges added for a new node in $V_{2}$. Hence,
the change of degree for $u$ is given by the weighted sum of these
two events:\begin{equation}
k_{u}(t+1)-k_{u}(t)=r_{1}\left[m_{11}\rho_{1}(k_{u})\right]+r_{2}\left[m_{12}\rho_{1}(k_{u})\right]\end{equation}
The probability $\rho_{1}(k_{v})$ is evolving with time and the denominator
depends on the the total degree of the partition $V_{1}$ given by:\begin{equation}
\sum_{\alpha\in V_{1}}k_{\alpha}\approx(2r_{1}m_{11}+m_{12})t\end{equation}
where the first term is the average degree contribution for a node
added to partition $V_{1}$ with probability $r_{1}$ and count $2m_{11}$
for each new node. The second term is the degree contribution from
the cross link since both partitions gain $m_{12}=m_{21}$ degree
for each new node. We can now adopt the continuum approach and write
the evolution equation for the degree $k_{u}$:\begin{equation}
\frac{\partial k_{u}(t)}{\partial t}=\left[\frac{r_{1}m_{11}+r_{2}m_{12}}{2r_{1}m_{11}+m_{12}}\right]\frac{k_{u}(t)}{t}\label{eq:scalefree-grow-continuum}\end{equation}
When a new node is added to the network at time $t_{i}$, it will
have $m_{1}=m_{11}+m_{12}$ degree initially. Hence, the initial condition
of the node $u$ is $k_{u}(t_{i})=m_{1}$ and so the evolving degree
is: \begin{equation}
k_{u}(t)=(t/t_{i})^{\frac{1}{\lambda_{1}}},\qquad\text{with }\lambda_{1}=\frac{2r_{1}m_{11}+m_{12}}{r_{1}m_{11}+r_{2}m_{12}}\end{equation}
The probability distribution of this partition $V_{1}$ is therefore
given by:\begin{equation}
\mathcal{P}_{1}(k)\sim\lambda_{1}m_{1}^{\lambda_{1}}k^{-\gamma_{1}}\end{equation}
where $\gamma_{1}=\lambda_{1}+1$ is the power of the tail of $\mathcal{P}_{1}(k)$.
Similar result can also be obtained for the partition $V_{2}$. Hence,
we obtain two degree distributions for two different partitions with
power $\gamma_{1}$ and $\gamma_{2}$ respectively. One can verify
that in the limiting case of $r_{2}=m_{12}=0$, the result reduced
to the BA model with $\gamma_{1}=3$. Also, if the internal partition
linkage $m_{ii}$ is strong for $i=1,2$, then $\gamma_{i}\approx3$
for both partitions which is similar to the BA model because of the
weak coupling between two partitions. By taking partial derivative
on $\gamma_{i}$ with respect to different variables, it can be noted
that the $\gamma_{i}$ is monotonic function for variables $r_{i}$
and $m_{ij}$ when $r_{i}\ne0.5$. The implication of this result
is that the power tail is $2<\gamma_{i}\le3$ for the region $0<r_{i}<0.5$
and $\gamma_{i}\ge3$ for the region $0.5<r_{i}<1$. Comparing two
degree distributions, the smaller partition has a slow decaying tail
while the larger partition has a fast decaying tail as shown in the
degree distribution in Fig. \ref{fig:scalefree}a.

Suppose we are now considering the degree distribution of the whole
network, the slow decaying tail of one partition can dominate the
high degree part so that only the tail with power $2<\gamma\le3$
can be observed, which is consistence with the measured power for
most world networks \cite{Albert2002-Rev}. In this model, $\gamma<3$
is the result of the existence of two classes of nodes and it implies
that, by simply measuring degree distribution, nodes with different
types cannot be distinguished. Hence, it is natural for us to ask
whether some real networks have multiple classes of nodes that are
yet to be unraveled. On the other hand, even though the degree distribution
is scale free, this model suggests that the total nearest neighbor
degree shows a different pattern for each partition.

From the above discussion, we know that the plotting of the total
neighbor degree and the node degree is scattered into two branches.
The result of a bipartite network without internal links is shown
in Fig. \ref{fig:scalefree}b. To find the neighbor degree distribution
and the slope $A_{i}$, one way is to use Eq. (\ref{eq:nei-deg-distri})
and (\ref{eq:nei-deg-slope}) by assuming the random connection: $p_{ii}=2m_{ii}/(Nr_{i})$
for $i=1,2$ and $p_{ij}=m_{ij}/(Nr_{i}r_{j})$ for $i\ne j$. For
this bipartite network, the resulting neighbor degree distribution
is good and the slope of total neighbor can be roughly approximated
by $A_{1}\approx\left\langle k^{2}\right\rangle _{2}/\left\langle k\right\rangle _{2}$
and $A_{2}\approx\left\langle k^{2}\right\rangle _{1}/\left\langle k\right\rangle _{1}$
as shown by two straight lines in the figure. Nonetheless, when there
are internal edges, the approximation of $A_{i}$ is not very good
because the model with internal degree correlation cannot be treated
as a simple random network.  As shown in the inset of Fig. \ref{fig:scalefree}b,
with existence of internal edges $m_{ii}=3$, the points fluctuate
more widely than the simple bipartite network. In this case, as expected,
the linear relation between degree and total neighbor degree for both
partitions are less fit and points are deviated more from the lines.
Therefore, the nodes for two partitions are partially mixed up and
less distinguishable from each other at the low degree part.

The distinctive separation between two sets of nodes in Fig. \ref{fig:scalefree}b,
especially at high degree or high total neighbor degree region, implies
that these nodes can be easily classified into two groups. This two
branches phenomenon does not occur for the BA model \cite{Ma&Szeto2006}
in which the local environment is homogeneous for all nodes with the
same degree. The points thus concentrate on a single line predicted
by AW law, which show the result similar either branch in Fig. \ref{fig:scalefree}b.
Hence, this phenomenon can be used to classify nodes into different
partitions as we are introduced in the next section.

\section{Degree-Neighbor degree correlation in real world network \label{sec:Degree-Neighbor-degree-correlation}}

With the two models discussed in the previous section, we know that
the total neighbor degree can be used to classify nodes. To test whether
the branching phenomenon exists in real world network, we examine
both explicitly bipartite and non-bipartite network. Undirected networks
are used in the simulation and the results are shown below.

\begin{figure}
\noindent \begin{centering}
\includegraphics[width=1\textwidth]{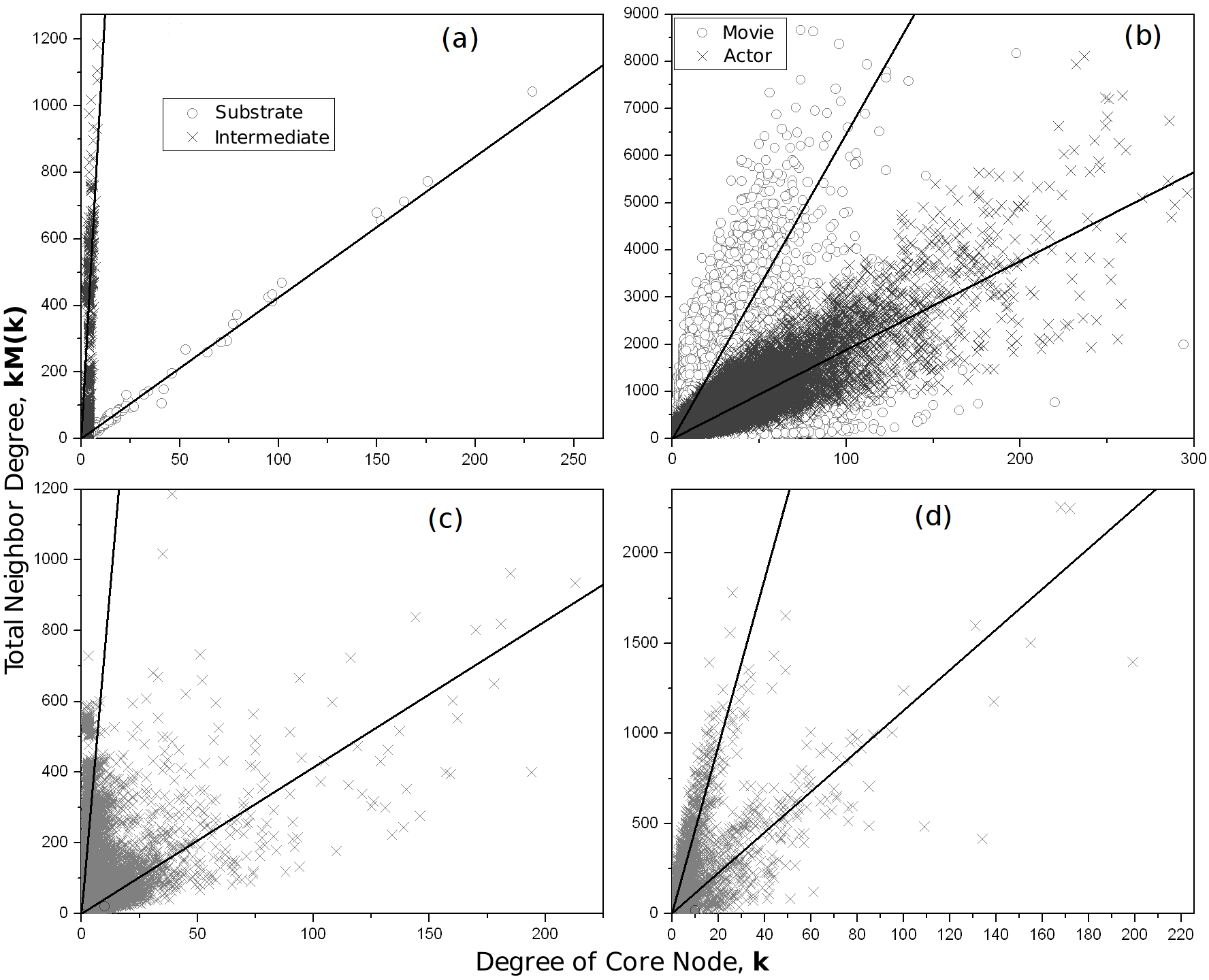}
\par\end{centering}

\caption{\label{fig:real} Scattered plot of total neighbor degree $kM(k)$
vs $k$, (a) the substrate and the intermediate in the metabolic network
(E. Coli), (b) movies and actors in actor network, (c) WordNet, (d)
California. The black lines are the fitting result of two branches.}

\end{figure}

The first example is metabolic networks \cite{metabolic-network}
which are explicitly bipartite. It is an interaction network composed
of the substrate and the intermediate complex. As shown in the Fig.
\ref{fig:real}a, the result of metabolic network is similar to the
preferential attachment model we introduced and two branches are clearly
identified. The substrate can have very high degree by its own nature
so the tail part of the degree distribution of the whole network is
dominated by power tail of substrate. In this case, the degree distribution
of intermediate nodes is shadowed by the substrate nodes. However,
in the scattered plot of the total neighbor degree, these two type
of nodes are clearly separated into two clusters. Another bipartite
network tested is the actor-movie network \cite{Barabasi1999-science,Albert1999}
in which edges represent a particular actor playing in a particular
movie. For this network, the degree distribution of both partitions
is close to each other. So, it is expected that the points of the
total neighbor degree for these two sets mixed at the region of low
degree as shown in the Fig. \ref{fig:real}b. Nevertheless, nodes
can still be distinguished clearly other than the low degree region.

Examples discuss above are explicitly bipartite so, in some sense,
they should be easily distinguishable. However, it is a challenge
to classified nodes into different groups for networks without having
any a prior knowledge on their origins. Now, similar method can be
employed to classify nodes by detecting the local inhomogeneity and
the branching in the total neighbor degree. One of the examples in
this category is the semantic network of the WordNet project \cite{Wordnet}
which studies the semantic relationship between different English
words. As shown in Fig. \ref{fig:real}c, two branches for this network
can be observed. Through the inspection of words in the network, it
can be concluded that the steeper branch contains words that are specialized
while the other branch corresponds to the generic words. Even though
specialized words have low degree, they can still have high total
neighbor degree because they are typically connected to generic words
that have high degree. Another one is the California web subgraph
networks \cite{Kleinberg1998,Kleinberg-database}. It is constructed
by linking webpages together depending on the querying results of
search engine. It is not an explicitly bipartite network, but two
different branches is clearly shown in the Fig. \ref{fig:real}d.

\begin{table}
\caption{\label{tab:fitting result} Results of two lines fitting for different
networks. $N_{1}$ and $N_{2}$ are the number of nodes classified
into the corresponding partitions by this method. $c$ is the ratio
of number of correctly classified nodes divided by total number nodes
in the network. }

\noindent \centering{}\begin{tabular}{cccccc}
\hline 
Network & $s_{1}$ & $s_{2}$ & $N_{1}$  & $N_{2}$ & $c$\tabularnewline
\hline
Random network in Fig. \ref{fig:random}a & 55.3 & 65.4 & 4038 & 5962 & 1.00\tabularnewline
Scale free network in Fig. \ref{fig:scalefree}b & 13.3 & 746 & 1942 & 8058 & 1.00\tabularnewline
Scale free network in inset of Fig. \ref{fig:scalefree}b  & 27.6 & 363 & 2013 & 7987 & 0.85\tabularnewline
Metabolic E. Coli \cite{metabolic-network} & 4.24 & 106 & 766 & 1509 & 0.92\tabularnewline
Actor-movie \cite{Barabasi1999-science} & 18.8 & 64.7 & 383640 & 127823 & 0.57\tabularnewline
WordNet network \cite{Wordnet} & 4.14 & 75.0 & 34652 & 42191 & -\tabularnewline
California subnetwork \cite{Kleinberg1998} & 11.2 & 46.4 & 2266 & 3909 & -\tabularnewline
\hline
\end{tabular}
\end{table}

To quantify the observation, we perform a least square fit to find
the best fitting lines and group nodes together. For a simple homogeneous
network, the Aboav-Weaire law predicts that the data point of average
total neighbor degree verse degree will fit into one single line.
Thus, for a network with two different partitions, we expect that
there should be two clear straight lines. With the same reason discussed
for the random $n$-partition network, the $y$-intercept is usually
small and we assume it to be zero. Hence, we look for the lines of
the form $y=s_{1}x$ and $y=s_{2}x$, with the slopes $s_{1}$ and
$s_{2}$ as the fitting parameters, such that the square of distance
between different points to the two lines is minimized:\begin{equation}
E(s_{1},s_{2})=\sum_{i=1}^{N}\min\left(\frac{(y_{i}-s_{1}x_{i})^{2}}{1+s_{1}^{2}},\frac{(y_{i}-s_{2}x_{i})^{2}}{1+s_{2}^{2}}\right)\end{equation}
Here, we can get the best fitting slopes $s_{1}$ and $s_{2}$ by
minimized $E(s_{1},s_{2})$. This method provides a simple classification
of nodes into two groups. If a network is homogeneous for the local
environment, then there may be only one group of nodes and the resulting
$s_{1}$ and $s_{2}$ should take a value close to each other. The
corresponding fitting results are plotted as two black straight lines
in the Fig. \ref{fig:real}. Moreover, the fitting and classification
results is shown in Table \ref{tab:fitting result} for networks used
in this paper. From the table, we can see that the classification
is very good for the models we studied, even for the preferential
attachment network with internal linkages. For the real world bipartite
networks, we can see that the classification is acceptable for the
metabolic network and the actor-movie network. For the non-bipartite
network, the fitting curves represent those two branches very good
in Fig. \ref{fig:real}. In addition, the large different in the values
of $s_{1}$ and $s_{2}$ signify that these networks are better be
described by two branches and so we can classify them into two groups.

\section{Summary \label{sec:Summary}}

In sum, we have studied nearest neighbor degree correlation for the
random multi-partition network, preferential attachment network and
some real world networks. Through the analysis of the extended AW
law, the exact neighbor degree distribution is computed for random
multi-partition network. Furthermore, we show that there is a linear
relationship between total neighbor degree and the node degree for
each partition separately, but not linear for the whole network. This
phenomenon is especially distinct for the preferential attachment
network which also model the scale free property with $2<\gamma\le3$.
The clustering of points in the scattered plot of total neighbor degree
verse degree therefore suggests a way to classify node into different
groups. By applying this classification scheme to the models studied
and real bipartite networks, we show that the grouping of node is
satisfactory. We also find an interesting subset of nodes in the WordNet
and California subgraph networks which are not bipartite.

\ack{}{}

K. Y. Szeto acknowledges the support of grant CERG 602506 and 602507.\bibliographystyle{unsrt}
\addcontentsline{toc}{section}{\refname}\bibliography{ref}

\end{document}